\documentclass[12pt]{article}

\usepackage{mathrsfs}
\usepackage[T1]{fontenc}
\usepackage{mathpazo}
\usepackage{setspace}
\usepackage{amsfonts}
\usepackage{amssymb}
\usepackage{amsmath}
\usepackage{epsfig}
\usepackage{latexsym}
\usepackage{color}
\usepackage{graphicx}
\usepackage{nicefrac}
\usepackage[latin1]{inputenc}
\usepackage{pstricks}
\usepackage{slashed}
\usepackage{multirow}

\def\hybrid{\topmargin -20pt    \oddsidemargin 0pt
        \headheight 0pt \headsep 0pt
        \textwidth 6.25in       
        \textheight 9.5in       
        \marginparwidth .875in
        \parskip 5pt plus 1pt   \jot = 1.5ex}

\hybrid

\def\baselinestretch{1.2}

\catcode`\@=11

\def\marginnote#1{}
\newcount\hour
\newcount\minute
\newtoks\amorpm
\hour=\time\divide\hour by60 \minute=\time{\multiply\hour by60
\global\advance\minute by-\hour}
\edef\standardtime{{\ifnum\hour<12 \global\amorpm={am}%
        \else\global\amorpm={pm}\advance\hour by-12 \fi
        \ifnum\hour=0 \hour=12 \fi
        \number\hour:\ifnum\minute<10 0\fi\number\minute\the\amorpm}}
\edef\militarytime{\number\hour:\ifnum\minute<10
0\fi\number\minute}

\def\draftlabel#1{{\@bsphack\if@filesw {\let\thepage\relax
   \xdef\@gtempa{\write\@auxout{\string
      \newlabel{#1}{{\@currentlabel}{\thepage}}}}}\@gtempa
   \if@nobreak \ifvmode\nobreak\fi\fi\fi\@esphack}
        \gdef\@eqnlabel{#1}}
\def\@eqnlabel{}
\def\@vacuum{}
\def\draftmarginnote#1{\marginpar{\raggedright\scriptsize\tt#1}}

\def\draft{\oddsidemargin -.5truein
        \def\@oddfoot{\sl preliminary draft \hfil
        \rm\thepage\hfil\sl\today\quad\militarytime}
        \let\@evenfoot\@oddfoot \overfullrule 3pt
        \let\label=\draftlabel
        \let\marginnote=\draftmarginnote
   \def\@eqnnum{(\theequation)\rlap{\kern\marginparsep\tt\@eqnlabel}%
\global\let\@eqnlabel\@vacuum}  }

\def\preprint{\twocolumn\sloppy\flushbottom\parindent 2em
        \leftmargini 2em\leftmarginv .5em\leftmarginvi .5em
        \oddsidemargin -.5in    \evensidemargin -.5in
        \columnsep .4in \footheight 0pt
        \textwidth 10.in        \topmargin  -.4in
        \headheight 12pt \topskip .4in
        \textheight 6.9in \footskip 0pt
        \def\@oddhead{\thepage\hfil\addtocounter{page}{1}\thepage}
        \let\@evenhead\@oddhead \def\@oddfoot{} \def\@evenfoot{} }

\def\numberbysection{\@addtoreset{equation}{section}
        \def\theequation{\thesection.\arabic{equation}}}

\def\underline#1{\relax\ifmmode\@@underline#1\else
        $\@@underline{\hbox{#1}}$\relax\fi}

\def\titlepage{\@restonecolfalse\if@twocolumn\@restonecoltrue\onecolumn
     \else \newpage \fi \thispagestyle{empty}\c@page\z@
        \def\thefootnote{\fnsymbol{footnote}} }

\def\endtitlepage{\if@restonecol\twocolumn \else \newpage \fi
        \def\thefootnote{\arabic{footnote}}
        \setcounter{footnote}{0}}  

\catcode`@=12 \relax

\def\figcap{\section*{Figure Captions\markboth
        {FIGURECAPTIONS}{FIGURECAPTIONS}}\list
        {Figure \arabic{enumi}:\hfill}{\settowidth\labelwidth{Figure
999:}
        \leftmargin\labelwidth
        \advance\leftmargin\labelsep\usecounter{enumi}}}
 \relax
\def\tablecap{\section*{Table Captions\markboth
        {TABLECAPTIONS}{TABLECAPTIONS}}\list
        {Table \arabic{enumi}:\hfill}{\settowidth\labelwidth{Table
999:}
        \leftmargin\labelwidth
        \advance\leftmargin\labelsep\usecounter{enumi}}}
 \relax
\def\reflist{\section*{References\markboth
        {REFLIST}{REFLIST}}\list
        {[\arabic{enumi}]\hfill}{\settowidth\labelwidth{[999]}
        \leftmargin\labelwidth
        \advance\leftmargin\labelsep\usecounter{enumi}}}
 \relax
\makeatletter
\newcounter{pubctr}
\def\publist{\@ifnextchar[{\@publist}{\@@publist}}
\def\@publist[#1]{\list
        {[\arabic{pubctr}]\hfill}{\settowidth\labelwidth{[999]}
        \leftmargin\labelwidth
        \advance\leftmargin\labelsep
        \@nmbrlisttrue\def\@listctr{pubctr}
        \setcounter{pubctr}{#1}\addtocounter{pubctr}{-1}}}
\def\@@publist{\list
        {[\arabic{pubctr}]\hfill}{\settowidth\labelwidth{[999]}
        \leftmargin\labelwidth
        \advance\leftmargin\labelsep
        \@nmbrlisttrue\def\@listctr{pubctr}}}
 \relax
\makeatother
\newskip\humongous \humongous=0pt plus 1000pt minus 1000pt

\newif\ifdtup

\relax

\def\be{\begin{equation}}
\def\ee{\end{equation}}
\def\ba{\begin{eqnarray}}
\def\ea{\end{eqnarray}}

\begin{document}

\renewcommand{\theequation}{\arabic{equation}}

\newcommand{\beq}{\begin{equation}}
\newcommand{\eeq}[1]{\label{#1}\end{equation}}
\newcommand{\ber}{\begin{eqnarray}}
\newcommand{\eer}[1]{\label{#1}\end{eqnarray}}
\newcommand{\eqn}[1]{(\ref{#1})}
\begin{titlepage}
\begin{center}

\vskip -.1 cm
\hfill April 2012\\

\vskip .4in

{\Large \bf Anomalies, instantons and chiral symmetry breaking}
\vskip.1cm
{\Large \bf at a Lifshitz point\footnote{Based on talks delivered at the
XLI-\`eme Institut d'\'et\'e de LPTENS {\em Cordes, Particules, et l'Univers},
17 August -- 2 September 2011, Paris, France and at the EISA Summer Institute
{\em Workshop on Fields and Strings: Theory -- Cosmology -- Phenomenology},
14 -- 18 September 2011, Corfu, Greece; contribution to appear in the Proceedings
of Science.}}

\vskip 0.8in

{\bf Ioannis Bakas} \vskip 0.2in
{\em Department of Physics, School of Applied Mathematics and Physical Sciences \\
National Technical University, 15780 Athens, Greece\\
\vskip 0.2in
\footnotesize{\tt bakas@mail.ntua.gr}}\\

\end{center}

\vskip 0.6in

\centerline{\bf Abstract}
\vskip 0.2in
\noindent
We give a new twist to an old-fashioned topic in quantum field theory describing
violations of the chiral charge conservation of massless fermions through
Adler-Bell-Jackiw anomalies in the background of instanton fields in the context
of non-relativistic Lifshitz theories. The results we report here summarize in
a nut-shell our earlier work on the subject found in arXiv:1103.5693 and
arXiv:1110.1332. We present simple examples where index computations can be carried 
out explicitly focusing, in particular, to gravitational models of Lifshitz type
and highlight their differences from ordinary gravity in four space-time dimensions.

\vfill
\end{titlepage}
\eject

\def\baselinestretch{1.2}
\baselineskip 16 pt \noindent

\setcounter{equation}{0}
\section{Introduction}

The axial anomalies arising upon quantization of massless fermions in a given gauge and/or
metric field background are deeply connected to the analytic index of the fermion operator.
The Atiyah-Singer index theorem asserts that the difference of positive and negative
chirality normalizable zero modes of the Dirac operator in a given background is provided
by (one-half) the integrated form of the anomalous axial current conservation law. This
profound relation constitutes the back-bone of our study and it can be easily established
without knowing the local form of the axial anomaly. It will be discussed first for
relativistic fermion theories in four space-time dimensions and then it will be generalized
to non-relativistic models of Lifshitz type. Of course, the computation of the index relies
on the form of the axial anomaly, which is a topological density given by the Chern-Pontryagin
class of the gauge and/or metric field background. The coefficient of the anomaly turns out to 
be universal (it is the same for both relativistic and Lifshitz theories) in accordance with the
general expectation that the axial anomaly is an infra-red phenomenon in disguise. Furthermore,
if the index of the Dirac operator is non-zero, the corresponding axial charge will not be
conserved in time, leading to violation of baryon and lepton number in physical processes 
within a given theory. These issues will be addressed in detail focusing on the similarities 
and the differences exhibited by relativistic and Lifshitz quantum field theories. 

The results we report in the following are restricted to four-dimensional theories and they 
are based on our previous work on the subject \cite{dieter, bakas} to which we refer the 
interested reader for further details. However, the presentation we adopt here is somewhat 
different emphasizing more the general ideas rather than the technical details. Also, the 
instanton backgrounds on which the index computations will be made are described in all 
generality in our work on gravitational Ho\v{r}ava-Lifshitz models \cite{BBLP, bakas2}.
References to other original papers can also be found there.

\section{Relativistic field theories}

First, to set up the stage, we consider the Dirac operator in a four-dimensional space-time
$M_4$ with Euclidean signature 
which may also be coupled to a (generally non-Abelian) background gauge field $A_{\mu}$ via
the rule of minimal substitution
\be
D_{\mu} = \partial _{\mu} + {1 \over 8} [\gamma_a , ~ \gamma_b] {\omega_{\mu}}^{ab} - i A_{\mu}
~.
\label{raloua}
\ee
Here, ${\omega_{\mu}}^{ab}$ are the components of the spin connection with tangent space-time 
indices $a$, $b$ and $\gamma_a$ are the corresponding Dirac matrices satisfying the 
anti-commutation relations 
\be
[\gamma_a , ~ \gamma_b]_+ = 2 \delta_{ab} ~. 
\ee
In most applications, the four-dimensional metric will be taken to be of the special 
form 
\be
ds^2 = dt^2 + g_{ij} (t, x) dx^i dx^j ~.
\label{loukos}
\ee
This is also appropriate for the description of Lifshitz theories satisfying the so called 
projectable condition, which will occupy most of our attention later. In the relativistic
case, however, the space-time metric can be of general form. 

Then, the Euclidean Dirac operator and the associated $\gamma_5$-matrix 
$\gamma_5 = - \gamma_0 \gamma_1 \gamma_2 \gamma_3$ that anti-commutes with it assume the 
following form in the chiral representation, respectively, 
\be
i \gamma^{\mu} D_{\mu} = i \left(\begin{array}{ccc}
0          &  & {\cal Q}_-\\
           &  &            \\
{\cal Q}_+ &  & 0
\end{array} \right), ~~~~~~
\gamma_5 = \left(\begin{array}{ccc}
\mathbb{1} &  & 0 \\
           &  &            \\
0          &  & -\mathbb{1}
\end{array} \right) ,
\ee
setting for notational convenience
\be
{\cal Q}_{\pm} = {\partial \over \partial t} \pm i \sigma_I {E_I}^i D_i ~.
\label{loura}
\ee
In writing \eqn{loura} we use the temporal choice $A_0 = 0$ in the presence of gauge
fields. We also use the Pauli matrices $\sigma_I$ as well as the inverse dreibeins ${E_I}^i$
associated to the metric $g_{ij}$ in \eqn{loukos} (summation over the space indices $i$ and
the tangent space indices $I$ are implicitly assumed). The $2 \times 2$ blocks 
${\cal Q}_{\pm}$ are first-order operators mapping the two-component Weyl spinors
$\Psi_{\pm}$ to $\Psi_{\mp}$ and they are mutually related by conjugation as 
$({\cal Q}_{\pm})^{\dagger} = - {\cal Q}_{\mp}$. The massless Dirac equation 
$i\gamma^{\mu} D_{\mu} \Psi (t, x) = 0$ acting on four-component spinors reduces to the 
following system of Weyl equations on $M_4$,
\be
{\cal Q}_{\pm} \Psi_{\pm} (t, x) = 0 ~,
\label{soutz}
\ee
whose number of normalizable solutions will be denoted by $n_{\pm}$, respectively.

The index of the Dirac operator is defined as the difference between the number of positive 
and negative chirality zero modes, i.e.,
\be
{\rm Ind} (D) = n_+ - n_- = {\rm dim ~ Ker} ~ {\cal Q}_+ - {\rm dim ~ Ker} ~ {\cal Q}_- =
{\rm Tr} ~ \gamma_5 ~.
\ee
The last equality is actually a tautology following from the chirality condition
$\gamma_5 \Psi_{\pm} = \pm \Psi_{\pm}$ by taking the trace over the zero energy states in
the fermion Hilbert space. One can extend the trace to the entire Hilbert space without
affecting the index because the non-zero energy states are always paired and for them 
the difference between the two chiralities cancel. Thus, only zero energy states
contribute to the index. Since the zero modes of ${\cal Q}_{\pm}$ are also zero modes of
${\cal Q}_{\mp} {\cal Q}_{\pm}$, the index takes the equivalent form
\be
{\rm Ind} (D) = {\rm dim ~ Ker} ~ ({\cal Q}_- {\cal Q}_+) - {\rm dim ~ Ker} ~
({\cal Q}_+ {\cal Q}_-) = {\rm Tr} \left(\gamma_5 ~ e^{- \tau (i \gamma^{\mu} D_{\mu})^2}
\right) .
\label{mourta}
\ee
The operators ${\cal Q}_{\mp} {\cal Q}_{\pm}$ are elliptic and they are better behaved
than ${\cal Q}_{\pm}$, since they are related to the square of the Dirac operator as  
\be
- (i \gamma^{\mu} D_{\mu})^2 = \left(\begin{array}{ccc}
{\cal Q}_- {\cal Q}_+   &  & 0\\
           &  &            \\
0 &  & {\cal Q}_+ {\cal Q}_-
\end{array} \right) . 
\ee
The equality \eqn{mourta} is very useful for comparison with the axial anomaly
computations. It holds for any $\tau > 0$ and
it provides a regulated version of ${\rm Tr} \gamma_5$. The trace is taken oven the
entire fermionic Hilbert space, since the non-zero energy states cancel each other.

The massless fermion theory with Lagrangian density ${\cal L} = \bar{\Psi} i \gamma^{\mu}
D_{\mu} \Psi$ exhibits an axial current conservation law $\nabla_{\mu} J_5^{\mu} (t, x) = 0$,
where $J_5^{\mu} (t, x) = \bar{\Psi} \gamma^{\mu} \gamma_5 \Psi$. There is an invariance of
the classical theory that can be easily found by applying Noether's procedure with respect
to the chiral rotations $\delta_{\epsilon} \Psi = i \epsilon \gamma_5 \Psi$. There is also
an associated chiral charge
\be
Q_5 = \int d^3 x \sqrt{{\rm det} g} ~ J_5^0 (t, x)
\ee
which is conserved in time under the appropriate boundary conditions at spatial infinity
(typically the spatial slices are taken to be compact without boundary). Quantum mechanically,
however, the situation changes drastically as there is an obstruction to the axial
current conservation law, called axial anomaly. The anomaly is most conveniently described
in the Euclidean domain after Wick rotation of the time coordinate and it originates from
the non-invariance of the fermionic path integral measure $({\cal D} \bar{\Psi})
({\cal D} \Psi)$ under chiral rotations. Careful investigation of the partition function
shows that the variation of the action combines with the variation of the measure to
produce the anomalous conservation law
\be
\nabla_{\mu} J_5^{\mu} (t, x) = 2 \lim_{\Lambda \rightarrow \infty} \sum_n
\varphi_n^{\dagger} (t, x) \gamma_5 ~ e^{- (i\gamma^{\mu} D_{\mu})^2 / \Lambda^2}
\varphi_n (t, x)
\label{tsoutsia}
\ee
after introducing a cut-off $\Lambda$ to regulate the infinite sum that otherwise is
ill-defined. Here, $\varphi_n (t, x)$ are the eigen-states of the interacting fermion
operator $i \gamma^{\mu} D_{\mu}$.

Integration over space-time is combined with the sum over $n$ in \eqn{tsoutsia} to
yield the trace over the entire fermionic Hilbert space of the theory.
Thus, comparison with the index formula \eqn{mourta} leads to the profound relation
\be
{\rm Ind} (D) = {1 \over 2} \int_{M_4} dt ~ d^3 x \sqrt{{\rm det} g} ~
\nabla_{\mu} J_5^{\mu} (t, x) ~,
\label{raka}
\ee
which connects the physics of axial anomalies with the mathematical theory
of Atiyah-Singer index theorem for the Dirac operator. Clearly, if the index is
non-zero, the axial charge $Q_5$ will not be conserved, leading to baryon and lepton 
number violation in the theory, as 
\be
\Delta Q_5 = 2 ~ {\rm Ind} (D) ~.
\ee

So far there has been no explicit reference to the form of the axial anomaly in the
background of gauge and/or metric fields. Likewise, there has been no reference to the
Atiyah-Singer formula for computing the index of the Dirac operator analytically. 
This step is crucial for telling the rest of the story in physics and mathematics. 
Manipulating the regulated sum \eqn{tsoutsia}, it turns out that
\be
\partial_{\mu} J_5^{\mu} = {1 \over 4 \pi^2} {\rm Tr}(F \wedge F) ~, ~~~~~~
\nabla_{\mu} J_5^{\mu} = {1 \over 96 \pi^2} {\rm Tr}(R \wedge R)
\label{expras}
\ee
for the gauge and the gravitational field contribution to the anomalous divergence
of the axial current, respectively. The obstruction to the axial current conservation
law can be easily guessed in both cases. The anomaly should be a topological density
built out of the field strength of the background fields, i.e., the curvature two-forms
$F$ and $R_{ab}$ of the gauge and metric fields, respectively, so that the divergence of
the axial current is a total derivative gauge invariant term. Furthermore, it should be 
odd under parity since $J_5^{\mu}$ is a pseudo-vector current. The only available such 
quantities in four space-time dimensions are the characteristic classes ${\rm Tr}(F \wedge F)$ 
and ${\rm Tr}(R \wedge R)$, where the trace is taken over the color indices of the
non-Abelian gauge field (per flavor of Dirac fermions) and the tangent space-time indices
of $M_4$, respectively. Then, it only remains to fix the coefficient of the anomaly to 
complete the derivation of \eqn{expras}. The outcome is the same using either physical or
mathematical techniques to make sense of the regulated sum for ${\rm Tr} \gamma_5$.

Soon after the discovery of instantons in four-dimensional gauge theories and gravity,
the question arose whether the index of the Dirac operator is non-zero on such topologically
non-trivial backgrounds. For non-Abelian gauge fields (e.g., $SU(2)$) on $M_4 \simeq S^4$, 
obtained as one-point compactification of $\mathbb{R}^4$ by imposing appropriate boundary 
conditions on $A_{\mu}(x)$, the instanton number $k \in \mathbb{Z}$ is provided by the second
Chern number
\be
k = {1 \over 8 \pi^2} \int_{M_4} {\rm Tr}(F \wedge F)
\ee
and, therefore, ${\rm Ind} (D) = k$ on such instanton backgrounds. Then, it follows that 
the axial charge conservation law is violated by $\Delta Q_5 = 2k$. Likewise, on compact
Riemannian manifolds without boundaries, but with non-vanishing Hirzebruch signature,
\be
\tau (M_4) = {1 \over 24 \pi^2} \int_{M_4} {\rm Tr}(R \wedge R) ~,
\ee
the index of the Dirac operator is non-zero, ${\rm Ind} (D) = \tau/8$ (recall that the
signature of all compact four dimensional spin manifolds without boundaries is integer
multiple of $8$). The only gravitational instanton of this kind is $K3$, which is a 
hyper-K\"ahler manifold with self-dual Riemann tensor whose signature is $16$ and
${\rm Ind} (D) = 2$.

In the presence of boundaries one considers bound states of the Dirac equation by
imposing appropriate boundary conditions on the spinors near $\partial M_4$ and uses the
Atiyah-Patodi-Singer (APS) index theorem to count the difference between positive and
negative chirality zero modes ($L^2$-index of the Dirac operator). The APS theorem is
based on the integrated form of the axial anomaly with the same bulk contribution as
before, but it also contains suitable boundary terms given by
\be
{\rm Ind}(D) = {1 \over 192 \pi^2} \int_{M_4} {\rm Tr} (R \wedge R) -
{1 \over 192 \pi^2} \int_{\partial M_{4}} {\rm Tr} (\theta \wedge R) - {1 \over 2}
\eta_{\rm D} (\partial M_4) ~.
\label{aps}
\ee
The first boundary term involves the Chern-Simons secondary characteristic class written
in terms of the second fundamental form $\theta$, which accounts for the possible deviation
of the space-time metric $ds^2 = dt^2 + g_{ij} (t, x) dx^i dx^j$ from cross-product form
at the boundary; as such it is a higher derivative analogue of the Gibbons-Hawking-York
boundary term encountered in general relativity. The second boundary term is non-local
and it is provided by the $\eta$-invariant of the tangential part of the Dirac operator
restricted to boundary; as such it counts the spectral asymmetry between positive and
negative chirality modes of the corresponding three-dimensional Dirac operator on
$\partial M_4$ and it is made rigorous using zeta-function regularization.

When $M_4 = I \times \Sigma_3$ with a compact three-manifold $\Sigma_3$ without boundaries
(e.g., $\Sigma_3 \simeq S^3$), the index of the Dirac operator on $M_4$ can be calculated
in practice by spectral flow methods. ${\rm Ind}(D)$ is provided by the net number of
level crossings that occur in the spectrum of the three-dimensional Dirac operator on
$\Sigma_3$,
\be
{\rm Ind}(D) = \Delta S(\Sigma_3) ~,
\label{specri}
\ee
as the metric $g_{ij} (t, x)$ deforms from one end of the time interval $I$ to 
the other; recall that the $\eta$-invariant jumps by $\pm 2$ units when an
eigen-value crosses from negative to positive values or conversely, and, therefore, each
level crossing contributes $\pm 1$ units to the index. Formula \eqn{specri} and its
generalization to Lifshitz theories is very useful for the applications. In those 
cases that the complete spectrum of the Dirac operator on $\Sigma_3$ can be found and 
the spectral flow can be studied explicitly, the index of the four-dimensional fermion 
operator on $I \times \Sigma_3$ can be computed in closed form.

It is well known that all non-compact instanton solutions of Einstein gravity can not 
support extreme geometric deformations on $\Sigma_3$ that are capable to induce level 
crossing, and, hence, ${\rm Ind}(D) = 0$. Prime examples of this kind are provided by 
the Taub-NUT and Eguchi-Hanson instantons which admit complete metrics \eqn{loukos} with
self-dual Riemann curvature tensor on $I \times \Sigma_3$ with $I$ being the semi-infinite 
real line of proper time $t$ and $\Sigma_3$ is $S^3$ and $S^3 / \mathbb{Z}_2$, 
respectively, endowed with homogeneous and partially isotropic geometries with 
$SU(2) \times U(1)$ isometry group. It can be explicitly seen in those cases that the 
individual terms contributing to the index \eqn{aps} cancel against each other, as
required on general grounds based on Lichnerowicz's theorem: non-compact four-metrics
with non-negative Ricci scalar curvature admit no bound state solutions of the Dirac
equation (if such states existed, they would be covariantly constant, and,
hence, non-normalizable leading to contradiction). Thus, it is not possible to have 
chiral symmetry breaking induced by gravitational instantons of topology 
$I \times \Sigma_3$ in ordinary Einstein-Dirac theory. 

\section{Lifshitz field theories}

Next, we extend the scope of our discussion to Lifshitz fermion theories with anisotropy
scaling exponent $z = 2 \alpha + 1$ by considering the non-relativistic analogue of the
Dirac operator
\be
i \gamma^{\mu} {\cal D}_{\mu} = i \gamma^0 D_0 + {1 \over 2} i \gamma^i
[D_i (- D_k D^k + M^2)^{\alpha} + (- D_k D^k + M^2)^{\alpha} D_i]
\label{arou}
\ee
in the presence of background gauge and/or metric fields. Here, $M$ is an arbitrary mass
scale that is introduced for convenience to extrapolate between the Dirac and
Lifshitz fermion models. The Lifshitz operator acts on four-component spinors $\Psi (t, x)$
and $D_0$ and $D_i$ are the time and space components of the ordinary covariant derivative
\eqn{raloua} that provides the minimal coupling to the background fields. Then, the massless
Lifshitz fermion theory in $3+1$ space-time dimensions is defined by the Lagrangian density
$\bar{\Psi} i \gamma^{\mu} {\cal D}_{\mu} \Psi$ and gives rise to the axial current conservation
law $\nabla_{\mu} J_5^{\mu} (t, x) = 0$, as in the relativistic case. The spatial components
of the axial current are different from the relativistic case, since they involve a number 
of derivatives that depend upon $z$, whereas the time component $J_5^0 (t, x)$, and, hence, the
axial charge $Q_5$ is the same. 

The axial symmetry is broken quantum mechanically and explicit computation shows that the 
anomalous term in the divergence of the axial current is identical to the relativistic case 
for the gauge and/or the metric field couplings, as in equation \eqn{expras}. More precisely,
the anomalous divergence of the axial current in the presence of gauge fields is 
\be
\partial_{\mu} J_5^{\mu} = {1 \over 4 \pi^2} \epsilon^{0ijk} {\rm Tr}(F_{0i} F_{jk}) = 
{1 \over 4 \pi^2} {\rm Tr}(F \wedge F) ~, 
\ee
whereas the metric field contribution to the axial anomaly turns out to be
\be
\nabla_{\mu} J_5^{\mu} = - {1 \over 96 \pi^2} \epsilon^{0ijk} {R^{ab}}_{0i} R_{ab ~ jk} = 
{1 \over 96 \pi^2} {\rm Tr}(R \wedge R) ~. 
\ee
Here, we write the result in $3+1$ terms using the electric and magnetic components of 
the corresponding curvature 2-forms and then recast it in the form \eqn{expras}. 
The calculation is performed in the Euclidean domain by analytic continuation of the
space-time foliation used for the formulation of Lifshitz models. It is also implicitly 
assumed that the space-time metric is restricted to the projectable case, meaning that 
the lapse and shift functions in the ADM decomposition of the metric are taken to be
$1$ and $0$, respectively, so that its analytic continuation is given by \eqn{loukos}
in proper time $x^0 = t$.

The arguments leading to the general relation \eqn{raka} between the index
of the fermion operator and the axial anomaly generalize easily to the Lifshitz operator 
\eqn{arou}. One simply has to consider the higher derivative analogue
of the operators ${\cal Q}_{\pm}$ shown in \eqn{loura} by replacing $D_i$ with
$[D_i (- D_k D^k + M^2)^{\alpha} + (- D_k D^k + M^2)^{\alpha} D_i]/2$ and also replace
the regulator ${\rm exp}[- (i\gamma^{\mu} D_{\mu})^2 / \Lambda^2]$ by
${\rm exp}[- (i\gamma^{\mu} {\cal D}_{\mu})^2 / \Lambda^{2z}]$ in the sum \eqn{tsoutsia}
which is now taken oven the eigen-states $\varphi_n (t, x)$ of the interacting Lifshitz
fermion operator $i\gamma^{\mu} {\cal D}_{\mu}$.
As a result, the index of the Lifshitz fermion operator \eqn{arou} is equal to the index 
of the Dirac operator for all values of the anisotropy exponent $z$,
\be
{\rm Ind}({\cal D}) = {\rm Ind}(D) ~.
\label{univad}
\ee

In space-times with boundaries the Atiyah-Patodi-Singer index theorem for the Lifshitz
fermion operator assumes the same form as in \eqn{aps}. The $\eta$-invariant is now
referring to the tangential part of the Lifshitz operator restricted to the boundary
$\partial M_4$, which turns out to be equal to the $\eta$-invariant of the corresponding
three-dimensional Dirac operator. Thus, relation \eqn{univad} is universal, as it
extends to all space-times with or without boundaries. If the index does not vanish, the
axial charge will not be conserved in time, i.e., $\Delta Q_5 = 2 ~ {\rm Ind}({\cal D})$,
leading to baryon and lepton number violation as before. It will be seen shortly that 
chiral symmetry breaking effects in non-relativistic theories of Lifshitz type are 
common practice.  

Anomalies, instantons and chiral symmetry breaking at a Lifshitz point are closely 
interrelated as in the relativistic case. The task is to find solutions of Lifshitz 
theories that allow for violation of chiral charge conservation and then compare the 
results to relativistic field theories. Here, we provide a brief account of bosonic
Lifshitz theories and review their instanton solutions, following our earlier work on the 
subject \cite{BBLP, bakas2}. In general, they provide classical backgrounds for Lifshitz 
fermion propagation in $3+1$ space-time dimensions. Special emphasis is placed on 
gravitational theories of Lifshitz type (in the so called Ho\v{r}ava-Lifshitz gravity) 
coupled to Lifshitz fermion models.  Then, ignoring the backreaction of fermions to the
gravitational instanton backgrounds, we find that -- unlike ordinary gravity -- 
non-conservation of $Q_5$ becomes possible for a certain range of the gravitational 
coupling parameters. We provide some simple solutions that realize this novel possibility 
and give a qualitative interpretation of its origin. The case of Lifshitz gauge
field theories appears to be conceptually simpler and looks easier for comparison with 
the relativistic case, but its instanton solutions are not explicitly known to this day; 
it is an open problem for future study which we hope to address elsewhere in detail.  

The bosonic Lifshitz field theories in $3+1$ space-time dimensions resemble point particle
systems with configuration space ${\cal C}$ and local coordinates $q_I$ that correspond
to Euclidean relativistic fields in three spatial dimensions. Thus, ${\cal C}$ is the
infinite dimensional space of all scalar, vector or more generally tensor fields on
a Riemannian manifold $\Sigma_3$, i.e., $q_I = \varphi (x)$, $A_i (x)$ or $g_{ij} (x)$
etc, which will be called superspace in all cases. These field theories are
non-relativistic models with anisotropic scaling in space and time $x \rightarrow ax$ and
$t \rightarrow a^z t$ with exponent $z$ which is provided by the order of the classical 
equations of motion of the relativistic fields $q_I$ defined on $\Sigma_3$.
Their action is often taken to be of the form
\be
S = {1 \over 2} \int dt \sum_{I, J} \left({d q_I \over dt} {\cal O}^{IJ} {d q_J \over dt}
- {\partial W \over \partial q_I} {\cal O}_{IJ} {\partial W \over \partial q_J} \right)
\label{parta}
\ee
assuming that the potential term is derivable from a superpotential $W[q]$ using the metric
${\cal O}^{IJ}$ and its inverse ${\cal O}_{IJ}$ on the superspace ${\cal C}$. This class
of Lifshitz models are said to satisfy the detailed balance condition with local 
superpotential functional $W$ being the action of a suitably chosen relativistic field 
theory on $\Sigma_3$. Of course, one may deviate from detailed balance by having additional 
terms in the potential that cannot be casted in the form \eqn{parta} using a local
superpotential functional, but such generalizations will not be in focus here. In all cases, 
the Lifshitz theories admit an effective point particle description in superspace that  
proves useful in many respects.

Next, we consider instanton solutions of Lifshitz theories on $\mathbb{R} \times \Sigma_3$,
assuming for simplicity that the metric in superspace is positive definite (it can also
become degenerate in some important cases that will be discussed later). We first note that
the minima of the potential provide static solutions of the equations of motion following
from \eqn{parta}; they are configurations that satisfy the classical equations of motion
$\partial W / \partial q_I = 0$ of the underlying relativistic field theory defined by the
action $W$ on $\Sigma_3$ and they are all degenerate with zero energy. Then, the instantons
are defined as extrema of the Euclidean action derived from \eqn{parta} by Wick rotation
$t \rightarrow it$ (equivalently by inverting the potential) that interpolate smoothly
between different degenerate vacua of the effective point particle system as $t$ extends
all the way from $-\infty$ to $+\infty$. By completing the square, as usual, it can be
easily seen that the instantons satisfy the system of first-order equations in time
\be
{d q_I \over dt} = \pm {\cal O}_{IJ} {\partial W \over \partial q_J}
\label{eterna}
\ee
and their action equals $S_{\rm instanton} = |\Delta W | \equiv |W(t = + \infty) -
W(t = - \infty)|$. The two choices of sign in \eqn{eterna} correspond to instanton and
anti-instanton configurations.

Thus, the instantons of Lifshitz theories (with detailed balance) are eternal solutions of
the gradient flow equations \eqn{eterna} derived from the superpotential functional $W$
and their action is finite, as required on general grounds. Yet explicit solutions of
the gradient flow equations can not be easily obtained unless additional symmetries are 
imposed on the fields, leading to consistent mini-superspace truncations of the 
configuration space ${\cal C}$. Otherwise, only qualitative features of the
solutions can be studied, in general, at least in those cases that the mathematical tools
of geometric analysis are powerful enough to explore the problem of long-time existence
against the possible formation of singularities along the flow lines. In gravitational 
theories of Lifshitz type the defining relations \eqn{eterna} are nothing else but 
geometric flows for the metrics on $\Sigma_3$. Specific examples and explicit solutions 
will be described later and compared to the instantons of Einstein gravity. Note 
for completeness that if we were considering Lifshitz theories without detailed balance,
instanton solutions would be much more difficult to find, if they existed at all
as finite Euclidean action configurations. 

Specializing to Ho\v{r}ava-Lifshitz gravity, we consider $(3+1)$-dimensional space-times
$M_4 \simeq \mathbb{R} \times \Sigma_3$ endowed with Lorentzian metrics
$ds^2 = -dt^2 + g_{ij}(t, x) dx^i dx^j$ (the so called projectable case) and write down the
following action in canonical form,
\be
S = {1 \over 2} \int_{M_4} dt d^3 x \sqrt{{\rm det} g} ~ \Big[
\left({\partial g_{ij} \over \partial t} \right)  {\cal G}^{ijkl}
\left({\partial g_{ij} \over \partial t}\right) - \left({1 \over 2 \sqrt{{\rm det} g}}
{\delta W \over \delta g_{ij}}\right) {\cal G}_{ijkl} \left({1 \over 2 \sqrt{{\rm det} g}}
{\delta W \over \delta g_{kl}}\right) \Big] ~,
\label{horav}
\ee
where
\be
{\cal G}^{ijkl} = {1 \over 2} (g^{ik} g^{jl} + g^{il} g^{jk}) - \lambda g^{ij} g^{kl}
\ee
is the ($\lambda$-deformed) DeWitt metric in superspace ${\cal C}$ consisting of all
Riemannian metrics on $\Sigma_3$ and ${\cal G}_{ijkl}$ is its inverse. We also choose as
superpotential functional $W[g]$ the action of topologically massive gravity on the
three-manifold $\Sigma_3$,
\be
W_{\rm TMG}[g] = {2 \over \kappa_{\rm w}^2} \int_{\Sigma_3} d^3x \sqrt{{\rm detg}} ~
(R-2\Lambda_{\rm w}) + {1 \over \omega} W_{\rm CS} [g] ~,
\ee
where
\be
W_{\rm CS} [g] = \int_{\Sigma_3} d^3x \sqrt{{\rm detg}} ~ \epsilon^{ijk} \Gamma_{im}^l
\left(\partial_j \Gamma_{lk}^m + {2 \over 3} \Gamma_{jn}^m \Gamma_{kl}^n \right)
\ee
is the gravitational Chern-Simons action, which is conveniently written here in terms of
the Christoffel symbols of the metric $g$ on $\Sigma_3$. The resulting non-relativistic
gravitational theory \eqn{horav} in $3+1$ dimensions exhibits anisotropic scaling $z=3$
which reduces to $z=2$ only when the gravitational Chern-Simons term is absent. 

The essential feature of topologically massive gravity is the presence of an adjustable 
scale that supplies the mass to one of the two helicity gravitons in the weak field 
approximation of the theory around flat space; the other helicity graviton remains massless 
and the two are interrelated by orientation reversing transformations on $\Sigma_3$ that 
flip the sign of $\omega$. The associated range of gravitational interactions in three 
dimensions is given by the relative ratio of the three-dimensional gravitational constant 
to the Chern-Simons coupling, $\kappa_{\rm w}^2 / |\omega|$, 
which will play role later in the formulation of a geometric criterion for having chiral 
symmetry breaking effects by gravitational instantons in the associated $(3+1)$-dimensional 
Ho\v{r}ava-Lifshitz gravity. Further generalizations arise by adding quadratic (or higher) 
curvature terms to $W$, as in three-dimensional new massive gravity (and generalizations 
thereof), which lead to $(3+1)$-dimensional Ho\v{r}ava-Lifshitz models with higher 
anisotropy scaling exponent $z$. In those cases, the three-dimensional gravitons remain  
massive (although the masses of the two helicity states need not be the same) leading to 
similar phenomena as with $W_{\rm TMG}$; such generalizations will not be addressed 
here at all to simplify the presentation.  

The instanton equation \eqn{eterna} specializes to the Ricci-Cotton flow. This is a third 
order equation derived from $W_{\rm TMG}[g]$ as gradient flow for the metric $g_{ij}$ on 
$\Sigma_3$. The flow lines depend on the superspace parameter $\lambda$, but the fixed 
points, which are classical solutions of three-dimensional topological massive gravity, 
do not depend on it for general values of $\lambda$. Here, we only consider the spacial 
case $\lambda = 1/3$, so that the defining equation of gravitational instantons takes 
the form 
\be
\partial_t g_{ij} = \mp {1 \over \kappa_{\rm w}^2} \left(R_{ij} - {1 \over 3} R g_{ij} 
\right) \pm {1 \over \omega} C_{ij} ~,
\label{ricot}
\ee
where $R_{ij}$ is the Ricci curvature tensor of $g$ and $C_{ij}$ its Cotton tensor. The 
driving curvature terms that arise in this case are traceless. This, in turn, implies that 
the fixed points have undetermined Ricci scalar curvature $R$, which can vary from one 
fixed point to the other. Said differently, we are considering a unimodular version of
three-dimensional topological massive gravity in which the cosmological constant 
$\Lambda_{\rm w}$ has the interpretation of an integration constant that can assume 
arbitrary values. This restriction is necessary for having level crossing, and, hence,  
$\Delta Q_5 \neq 0$ as $t$ varies from $-\infty$ to $+\infty$ in the corresponding 
gravitational instanton background. Otherwise, if $\lambda \neq 1/3$, the two end-points 
of the instanton will have the same curvature $R (= 6 \Lambda_{\rm w})$ forbidding any 
net level crossing to occur in the theory. 

More technically speaking, the metric in superspace takes the following form at 
$\lambda = 1/3$,  
\be
{\cal G}^{ijkl} = {1 \over 2} (g^{ik} g^{jl} + g^{il} g^{jk}) - {1 \over 3} g^{ij} g^{kl} ~, 
\ee
and as such it projects any symmetric two-tensor to its traceless part. Then, the inverse 
metric in superspace becomes ill-defined. Instead, one defines 
\be
{\cal G}_{ijkl} = {1 \over 2} (g_{ik} g_{jl} + g_{il} g_{jk}) - {1 \over 3} g_{ij} g_{kl} ~,
\ee
which follows formally from the inverse metric as $\lambda \rightarrow \pm \infty$ and it 
also projects any symmetric two-tensor to its traceless part. The two quantities are simply 
related to each other by the generalized orthonormality condition
\be
{\cal G}^{ijkl} {\cal G}_{klmn} = {1 \over 2} \left(\delta_m^i \delta_n^j + 
\delta_n^i \delta_m^j \right) - {1 \over 3} g^{ij} g_{mn} 
\ee
that follows from the standard one by subtracting the trace part for consistency of the 
projection. As a result, the conformal factor of the metric decouples from the dynamics
and the Ricci-Cotton flow \eqn{ricot} preserves the volume of space ${\rm Vol} (\Sigma^3)$.
This is precisely the class of models that allow for violations of chiral charge 
conservation by gravitational instanton effects in Ho\v{r}ava-Lifshitz gravity. 

Our next task is to consider simple examples that illustrate the situation and derive 
the necessary and sufficient conditions on the parameters of the theory for having chiral 
symmetry breaking. We consider the simple case of Bianchi IX homogeneous geometries on
$\Sigma_3 = S^3$ with isometry group $SU(2) \times U(1)$, known as Berger spheres,
\be
ds^2 = \gamma (t) \Big[ (\sigma^1)^2 + (\sigma^2)^2 + \delta^2 (t) (\sigma^3)^2 \Big] ~,
\ee
which provide consistent mini-superspace reduction of the Ricci-Cotton flow to an ordinary 
differential equation. Here, $\sigma^I$ are the left-invariant 1-forms of $SU(2)$ 
satisfying 
\be
d\sigma^I + {1 \over 2} {\epsilon^I}_{JK} \sigma^J \wedge \sigma^K = 0 
\ee
and $\delta \in [0, ~ \infty)$ is a parameter measuring the anisotropy of the model. 
Instanton solutions of Ho\v{r}ava-Lifshitz gravity can be explicitly constructed in this
case and then compared to the analogous solutions (Eguchi-Hanson and Taub-NUT) of 
Euclidean Einstein gravity. Homogeneous solutions with higher degree of anisotropy 
can also be studied by relaxing the additional $U(1)$ isometry (axial symmetry) of the 
3-sphere, but they will not be discussed here as they introduce unnecessary technical 
complications. 

The normalized Ricci-Cotton flow \eqn{ricot} admits the round metric with $\delta = 1$ 
as fixed point, but it also admits a second fixed point within the class of Berger 
sphere metrics provided that $\omega < 0$  (for a given choice of orientation of $S^3$) 
with 
\be
{1 \over 3} + {\kappa_{\rm w}^2 \delta \over 2 \omega \sqrt{\gamma}} = 0 ~. 
\label{othfix}
\ee
The presence of two fixed points is prerequisite for the existence of instantons. There 
is a smooth flow line interpolating between these two fixed points, but the precise form
of the solution will not be important in the following. All it matters here is the 
variation of the 3-curvature 
\be
R = {1 \over 2 \gamma} (4- \delta^2) 
\label{kourva}
\ee
as one moves from one fixed point to the other and the ability to induce level crossing 
by changing the shape of $S^3$, whereas the volume 
${\rm Vol} (S^3) = 16 \pi^2 \delta \gamma^{3/2}$ remains fixed throughout the evolution. 
Fortunately, the computation of the index of the Dirac-Lifshitz operator is a 
tractable problem in the background of gravitational instanton solutions with 
$SU(2) \times U(1)$ isometry. 

Let $\zeta$ denote the eigen-values of the three-dimensional Dirac operator 
$i \gamma^i D_i$ on a Berger sphere, which can be determined in closed form together 
with their multiplicities. They split into positive and negative eigen-values that 
depend upon $\delta$. Zero modes (often called harmonic spinors) also arise for 
special values of $\delta \geq 4$,  
\be
\delta^2 = 2 \sqrt{4pq \delta^2 + (p-q)^2} ~, ~~~~~~ (p, q) \in \mathbb{N}^2  
\label{moutsia}
\ee
with multiplicities $p+q$. Setting $p=q=1$ it follows that the first zero modes of 
$i \gamma^i D_i$ arise when $\delta = 4$, which is the critical value of the 
anisotropy parameter for inducing level crossing by varying $\delta$. Comparison 
with equation \eqn{kourva} shows that the curvature of the Berger sphere should be
sufficiently negative to allow for the occurrence of zero modes and subsequently for 
level crossing as $\delta > 4$. This is also consistent with Lichnerowicz's theorem 
for the three-dimensional Dirac operator that requires negative curvature for the 
existence of harmonic spinors.  

Extending the discussion to the three-dimensional Dirac-Lifshitz operator 
$i \gamma^i {\cal D}_i$, we note that its eigen-values $Z$ on Berger spheres are simply 
expressed in terms of the eigen-values $\zeta$ of the corresponding Dirac operator as  
\be
Z = \zeta \left(\zeta^2 + {1 \over 8 \gamma} (\delta^2 - 4) + M^2 \right)^{\alpha} . 
\ee
It turns out that $Z$ have the same sign as $\zeta$ for all values of $\delta$, and, 
hence, level crossing occurs at the values of $\delta$ given by \eqn{moutsia} as before.
Also, since the multiplicity of the eigen-values of $Z$ is the same as $\zeta$, the 
number of modes that undergoes level crossing, $\Delta S(S^3)$,  is the same for both 
operators. This is consistent with the fact that the index of the four-dimensional 
Dirac-Lifshitz operator is the same as the index of the Dirac operator on all such 
geometrical backgrounds $\mathbb{R} \times S^3$, i.e., ${\rm Ind}({\cal D})={\rm Ind}(D)$, 
and which can be computed by spectral flow methods via equation \eqn{specri}.  

The instanton of $z=3$ Ho\v{r}ava-Lifshitz gravity with $SU(2) \times U(1)$ isometry 
that interpolates smoothly between the round sphere and the Berger sphere with 
$\delta$ given by \eqn{othfix} leads to violation of chiral charge conservation provided
that the volume of space (which remains fixed for all $t$) is bounded from below as
\be
{\rm Vol} (S^3) = 16 \pi^2 \delta \gamma^{3/2} = 
54 \pi^2 \delta^4 \left(- {\kappa_{\rm w}^2 \over \omega} \right)^3 > 
13824 \pi^2 \left(- {\kappa_{\rm w}^2 \over \omega} \right)^3 
\ee
letting $\delta > 4$ in the final step. Thus, chiral symmetry breaking becomes possible 
when the mean radius of space is sufficiently larger than the range of interaction 
$\sim \kappa_{\rm w}^2 / |\omega|$ in topologically massive gravity, which is associated 
to the superpotential functional $W[g]$. This provides a qualitative criterion for chiral 
symmetry breaking in Ho\v{r}ava-Lifshitz gravity that satisfies the detailed balance 
condition and which generalizes beyond the simple example we have considered here. 
$\Delta Q_5$ is given by the number of modes (including their multiplicities) that have 
undergone level crossing. 

Comparison to Einstein gravity reveals that novel phenomena become possible in 
non-relativistic gravitational theories leading to fermion number violation. According 
to a scenario, such theories are thought to provide the ultra-violet completion of gravity 
sacrificing relativistic invariance for power counting renormalizability at very high 
energies. The phase of Ho\v{r}ava-Lifshitz gravity arising at $\lambda = 1/3$ 
seems to be more appropriate in this context although the flow to
ordinary gravity in the infra-red regime and the emergence of relativistic invariance as 
low energy phenomenon has not been made quantitative to this day. The asymptotic safety 
programme for gravity seems to provide a promising framework to address this fundamental 
issue in future studies.

\section{Conclusions}

We outlined the general relation between anomalies, instantons and chiral symmetry 
breaking in relativistic and Lifshitz field theories. The quantum anomaly of the axial 
current conservation law of massless fermions is independent of the anisotropy scaling 
parameter and coincides with the result obtained for the relativistic case in the 
background of gauge and metric fields. Likewise, the index of the Dirac-Lifshitz 
operator is universal given by the integrated form of the axial anomaly. This is 
in agreement with the infra-red nature of the axial anomaly, which is inert to 
higher derivative terms that become relevant in the ultra-violet regime. 

The difference between relativistic and Lifshitz field theories lies in the ability  
of their instantons to affect the conservation law of chiral charge. 
The main result in this context is the construction of simple instanton
solutions of gravitational Lifshitz theories and their use to induce chiral symmetry 
breaking for certain range of the couplings, leading to baryon and lepton number 
violation triggered by gravity. This novel possibility does not arise in general 
relativity. It remains to construct instanton solutions of Lifshitz gauge theories 
and examine their effect on chiral symmetry breaking in comparison to ordinary 
gauge theories.    

\vskip1cm
\centerline{\bf \large Acknowledgements} 
\noindent
I thank the conference organizers for their kind invitation to give an account of this
work in an exciting scientific environment as well as for the partial financial 
support.

\end{document}